\documentclass[a4paper,12pt]{article}
\usepackage{lipsum}
\usepackage{amsmath,amssymb,amsfonts}
\usepackage[pdftex]{graphicx,color}
\usepackage{amsthm,epsfig,epstopdf,url,array}
\usepackage{mathtools,amsthm}
\usepackage{enumerate}
\usepackage{graphicx}
\theoremstyle{plain}
\newtheorem{thm}{Theorem}[section]
\newtheorem{lem}{Lemma}[section]
\newtheorem{prop}{Proposition}[section]

\theoremstyle{definition}
\newtheorem{defn}{Definition}[section]

\theoremstyle{remark}
\newtheorem{rem:}{Remark}
\newtheorem{note}{Note}

\usepackage[a4paper,includeheadfoot,margin=2.54cm]{geometry}

\begin{document}

\title{Bitopological Duality for Algebras of Fitting's logic and Natural Duality extension}
\author {Litan Kumar Das\thanks{E-mail:ld06iitkgp@gmail.com}
 \and  Kumar Sankar Ray\thanks{E-mail:ksray@isical.ac.in}
}

 \maketitle

{\bf Abstract.} 
                  In this paper, we investigate a bitopological duality for algebras of Fitting's multi-valued  logic. We also extend the natural duality theory for $\mathbb{ISP_I}(\mathcal{L})$ by developing a duality for $\mathbb{ISP}(\mathcal{L})$, where $\mathcal{L}$ is a finite algebra in which underlying lattice is bounded distributive.\\

\section{Introduction}
Stone's pioneering work in the mid 1930 \cite{stone1938representation} on the dual equivalence between the category of Boolean algebras and homomorphism, and the category of Stone spaces(compact zero-dimensional Hausdorff spaces) and continuous maps, is being considered as the origin of duality theory. Stone further developed a general work \cite{stone1937topological} for the category of bounded distributive lattices in 1937. Priestley in 1970 \cite{priestley1970representation} investigate another duality for the category of bounded distributive lattices with the help of ordered Stone spaces(known as Priesley spaces), which overcome difficulties in Stone's work \cite{stone1937topological}. Priestley duality was appiled extensively by the logicians. A duality for Heyting algebras was worked out by Esakia \cite{esakia2019heyting}, which is a restriction of Priestley duality. In logical point of view, topological duality have been appiled for a conection between syntactic and semantic. Several authors have benn developed duality from differnt angle (e.g.,\cite{maruyama2011dualities,ray2018categorical,teheux2007duality,hansoul1983duality}).\\
In \cite{jung2006bitopological}, Jung and Moshier illustrated an idea to represent distributive lattices using  bitopological spaces. As a result, an alternative description of Esakia duality in bitopological setting was revealed by the authors in \cite{bezhanishvili2010bitopological}. Besides, Lauridsen in \cite{lauridsen2015bitopological} demonstrated the notion of bitopological Vietoris space and developed coalgebraic semantics for positive modal logic. Lauridsen's work leads us to think categorical duality for an algebraic counterpart of Fitting's lattice-valued logic \cite{maruyama2009algebraic,fitting1991many,fitting1992many} in the bi-topological setting.\\
The natural duality theory was developed by Davey and Clark in \cite{clark1998natural}, is a powerfull tool  based on universal algebra. They work mainly to develop a duality for $\mathbb{ISP}(\mathcal{L})$, where $\mathcal{L}$ is taken as a finite algebra. It has been studied extensively by mathematicians, logisticians as well as computer scientists to achieve a new dualism. By means of Hu-duality \cite{hu1971topological}, Maruyamma in \cite{maruyama2012natural} develop a duality for $\mathbb{ISP_{M}}(\mathcal{L})$, which is a modalization of the notoin of $\mathbb{ISP}(\mathcal{L})$, and succesfully unified J{\'o}nsson-Tarski duality and Abramsky-Kupke-Kurz-Venema duality. In addition to this, He also suggested the ideas $\mathbb{ISP_{I}}(\mathcal{L})$ and $\mathbb{IS_RP(\mathcal{L})}$, which were the two different point of view on intutionistic logic. In \cite{maruyamauniversal}, Maruyama obtained a duality for $\mathbb{IS_RP(\mathcal{L})}$ and incorporate Esakia duality in natural duality theory. In our work, we develop a duality for $\mathbb{ISP_{I}}(\mathcal{L})$.\\
The paper is arranged as follows.\\
In Section \ref{sec:1}, we have mentioned some basic concepts that are required for this work. We define a category $PBS_{\mathcal{L}}$ in Section \ref{sec:2}, and then establish a categorical duality with the category $\mathcal{VA_{\mathcal{L}}}$ of algebras of Fitting's lattice-valued logic, and homomorphism. In Section \ref{sec:3}, we define a category $ISP_I(\mathcal{L})$ and show the duality by developing a duality for the category $ISP(\mathcal{L})$. The paper is concluded in Section \ref{sec:4} by mentioning future directions of work.

\section{Preliminaries}
\label{sec:1}
For category theory, we refer to \cite{adamek1990h,mac2013categories}. We cite \cite{davey2002introduction,sankappanavar1981course} for lattice theory and universal algebra. For the sake of clarity and understanding of our work, we consider some useful concepts.\\
An algebra $\mathcal{L}$-$\mathcal{VL}$ of Fitting's finite distributive lattice$(\mathcal{L})$-valued logic was revealed by Maruyama in \cite{maruyama2009algebraic} with out considering fuzzy-truth constans. This modification was done by taking a new connective $T_{\ell}(-)$, $\ell\in\mathcal{L}$. Details about the connective $T_{\ell}(-)$ and its origin given in \cite{maruyama2009algebraic,epstein1960lattice}.
\begin{defn}[\cite{maruyama2009algebraic}]
	\label{S1}
	$(\mathcal{A},\wedge,\vee,\rightarrow,T_L(L\in\mathcal{L}), 0,1)$ forms a $\mathcal{L}$-$\mathcal{VL}$-algebra if and only if for any $L_1, L_2\in \mathcal{L}$, and $a, b\in \mathcal{A}$ the following conditions hold:
	\begin{enumerate}[(i)]
		\item the algebraic structure $(\mathcal{A},\wedge,\vee,\rightarrow,0,1)$ is a Heyting algebra;
	
		\item $T_{L_1}(a)\wedge T_{L_2}(b)\leq T_{L_1\rightarrow L_2}(a\rightarrow b)\wedge T_{L_1\wedge L_2}(a\wedge b)\wedge T_{L_1\vee L_2}(a\vee b)$;\\
		$T_{L_2}(a)\leq T_{T_{L_1}(L_2)}(T_{L_1}(a))$;
		\item $T_0(0)=1$; $T_{L}(0)=0$ ($L\neq 0$), $T_1(1)=1$, $T_L(1)=0, L\neq 1$;
		\item $\bigvee\{T_L(a): L\in\mathcal{L}\}=1$, $T_{L_1}(a)\vee (T_{L_2}(a)\rightarrow 0)=1$;\\
		$T_{L_1}(a)\wedge T_{L_2}(a)=0$ ($L_1\neq L_2$);
		\item $T_1(T_L(a))=T_L(a)$, $T_0(T_L(a))=T_L(a)\rightarrow 0$, $T_{L_2}(T_{L_1}(a))=0$ ($L_2\neq 0,1$);
		\item $T_1(a)\leq a$, $T_1(a\wedge b)=T_1(a)\wedge T_1(b)$;
		\item $\displaystyle\bigwedge_{L\in\mathcal{L}}(T_L(a)\leftrightarrow T_L(b))\leq (a\leftrightarrow b)$.
	\end{enumerate}
	
\end{defn}
\begin{defn}[\cite{salbany1974bitopological}]
	\label{DS2}
	A bitopological space $(\mathcal{S},\tau_1^{\mathcal{S}},\tau_2^{\mathcal{S}})$ is said to be pairwise Hausdorff if for any two different points $s_1, s_2$ of $\mathcal{S}$ there exist a disjoint open sets $\mathcal{O}_1\in \tau_1^{\mathcal{S}}$, and  
	$\mathcal{O}_2\in \tau_2^{\mathcal{S}}$ containing $s_1$ and $s_2$, respectively.
\end{defn}
\begin{defn}[\cite{salbany1974bitopological}]
	\label{DS3}
	A bitopological space $(\mathcal{S},\tau_1^{\mathcal{S}},\tau_2^{\mathcal{S}})$ is said to be pairwise zero-dimensional if the collection $B_1^{\mathcal{S}}$ of $\tau_1^{\mathcal{S}}$-open sets which are $\tau_2^{\mathcal{S}}$-closed is a basis for the topology $\tau_1^{\mathcal{S}}$, and the collection $B_2^{\mathcal{S}}$  of $\tau_2^{\mathcal{S}}$-open sets which are $\tau_1^{\mathcal{S}}$-closed is a basis for the topology $\tau_2^{\mathcal{S}}$, i.e., we can write $B_1^{\mathcal{S}}=\tau_1^{\mathcal{S}}\cap \varrho_2$, and $B_2^{\mathcal{S}}=\tau_2^{\mathcal{S}}\cap \varrho_1$. Here we designate $\varrho_1$, and $\varrho_2$ as the collections of $\tau_1^{\mathcal{S}}$-closed sets and $\tau_2^{\mathcal{S}}$-closed sets, respectively.
\end{defn}  
\begin{defn}[\cite{salbany1974bitopological}]
	\label{DS4}
	A bitopological space $(\mathcal{S},\tau_1^{\mathcal{S}},\tau_2^{\mathcal{S}})$ is said to be pairwise compact if every open cover $\{\mathcal{O}_i :i\in J, \mathcal{O}_i\in \tau_1^{\mathcal{S}}\cup \tau_2^{\mathcal{S}}\}$ of $\mathcal{S}$ has a finite subcover.
\end{defn}    
\begin{prop}[\cite{bezhanishvili2010bitopological}]
	\label{PS1}
	A bitopological space $(\mathcal{S},\tau_1^{\mathcal{S}},\tau_2^{\mathcal{S}})$ is pairwise compact if and only if $\varrho_1\subset\Upsilon_2$ and $\varrho_2\subset\Upsilon_1$, where $\Upsilon_1$, and $\Upsilon_2$ are denote respectively the set of all compact subsets of $(\mathcal{S},\tau_1^{\mathcal{S}})$ and $(\mathcal{S},\tau_2^{\mathcal{S}})$.
\end{prop}

\section{Bitopological duality for $\mathcal{L}$-$\mathcal{VL}$ Algebras:  }
\label{sec:2}
Throughout this section, $\mathcal{L}$ is taken as a bounded finite distributive lattice. Henceforth, $\mathcal{L}$ is a Heyting algebra.

For a pairwise Boolean space $\mathcal{B}$, we denote the set of all pairwise closed sub-spaces of $\mathcal{B}$ by $\Omega_{\mathcal{B}}$. 
As a pairwise closed subset of a pairwise compact space is also a pairwise compact \cite{Adnadzhevich1987bicompact}, so each member of $\Omega_{\mathcal{B}}$ is a pairwise Boolean space. We mean a pairwise Boolean space as a pairwise zero-dimensional, pairwise Hausdorff, and a pairwise compact space. We take $\Sigma_{\mathcal{L}}$ to denote the collections of subalgebras of $\mathcal{L}$.

\begin{defn}
\label{D4S}
We define the category $PBS_{\mathcal{L}}$ as follows:
\begin{enumerate}
\item An object in $PBS_{\mathcal{L}}$ is taken as $(\mathcal{B},\alpha_{\mathcal{B}})$, where $\mathcal{B}$ is a pairwise Boolean space, and $\alpha_{\mathcal{B}}$ is a mapping from $\Sigma_{\mathcal{L}}$ to  $\Sigma_{\mathcal{B}}$ satisfying the following rule:
\begin{enumerate}[(i)]
\item $\alpha_{\mathcal{B}}(\mathcal{L})=\mathcal{B}$;
\item for any $\mathcal{L}_1, \mathcal{L}_2,\mathcal{L}_3\in \Sigma_{\mathcal{L}}$, if $\mathcal{L}_1=\mathcal{L}_2\wedge \mathcal{L}_3$, then $\alpha_{\mathcal{B}}(\mathcal{L}_1)=\alpha_{\mathcal{B}}(\mathcal{L}_2)\cap \alpha_{\mathcal{B}}(\mathcal{L}_3)$.
\end{enumerate}
\item A morphism $f: (\mathcal{B},\alpha_{\mathcal{B}})\rightarrow (\mathcal{B}',\alpha_{\mathcal{B}'})$ in $PBS_{\mathcal{L}}$ is a pairwise continuous map $f : \mathcal{B}\rightarrow \mathcal{B}'$ which satisfies;
\begin{enumerate}[(i)]
\item if $x\in \alpha_{\mathcal{B}}(\mathcal{L'})$, $\mathcal{L'}\in \Sigma_{\mathcal{L}}$, then $f(x)\in \alpha_{\mathcal{B'}}(\mathcal{L'})$
\end{enumerate}

\end{enumerate}
\end{defn}
We consider a bitopological space $(\mathcal{L},\tau,\tau)$, where $\tau$ is the discrete topology, and consequently, $(\mathcal{L},\alpha_{\mathcal{L}})$, 
where the mapping $\alpha_{\mathcal{L}} :\Sigma_{\mathcal{L}}\rightarrow \Omega_{\mathcal{L}}$ is defined by $\alpha_{\mathcal{L}}(\mathcal{L}')=\mathcal{L}'$, 
is an object of the category $PBS_{\mathcal{L}}$. For a $\mathcal{L}$-$\mathcal{VL}$ algebra $\mathcal{A}$, 
consider a bitopological space $(Hom_{\mathcal{VA}_{\mathcal{L}}}(\mathcal{A},\mathcal{L}),\sigma_1,\sigma_2)$.
 Let $B^{\sigma_1}$ and  $B^{\sigma_2}$ be the bases for the topologies $\sigma_1$, $\sigma_2$ respectively,
 where  $B^{\sigma_1}=\{\langle a \rangle | a\in \mathcal{A}\}$, $\langle a \rangle=\{v\in Hom_{\mathcal{VA}_{\mathcal{L}}}(\mathcal{A},\mathcal{L}) : v(a)=1\}$, and $B^{\sigma_2}=\{ \mathcal{O}^c :\mathcal{O}\in B^{\sigma_1}\}$.

\begin{lem}
\label{L1}
The bitopological space $(Hom_{\mathcal{VA}_{\mathcal{L}}}(\mathcal{A},\mathcal{L}),\sigma_1,\sigma_2)$ is a pairwise Boolean space.
\begin{proof}
We first show that $(Hom_{\mathcal{VA}_{\mathcal{L}}}(\mathcal{A},\mathcal{L}),\sigma_1,\sigma_2)$ is a pairwise Hausdorff. Let $v_1, v_2 \in (Hom_{\mathcal{VA}_{\mathcal{L}}}(\mathcal{A},\mathcal{L})$
 such that $v_1\neq v_2$. Then there exists an element $a \in \mathcal{A}$ such that $v_1\in \langle a \rangle$, but $v_2 \notin \langle a \rangle$, and thus there exists disjoint open sets $U\in \sigma_1$, $V\in \sigma_2$ such that $v_1\in U$, and $v_2\in V$.\\
We next show that $(Hom_{\mathcal{VA}_{\mathcal{L}}}(\mathcal{A},\mathcal{L}),\sigma_1,\sigma_2)$ is pairwise compact. Here we observed that $\sigma_1\cup\sigma_2 \subset \sigma_1$.
Since, $(Hom_{\mathcal{VA}_{\mathcal{L}}}(\mathcal{A},\mathcal{L}),\sigma_1)$ is compact, it follows that
 $(Hom_{\mathcal{VA}_{\mathcal{L}}}(\mathcal{A},\mathcal{L}),\sigma_1,\sigma_2)$ is pairwise compact.\\
 To prove that $(Hom_{\mathcal{VA}_{\mathcal{L}}}(\mathcal{A},\mathcal{L}),\sigma_1,\sigma_2)$ is pairwise zero-dimensional, we shall show that $B^{\sigma_1}=\sigma_1\cap \varrho_2$, 
 and $B^{\sigma_2}=\sigma_2\cap \varrho_1$. We observe that if $u\in B^{\sigma_1}$, then $u\in \sigma_1$. Since $u\in B^{\sigma_1}$, we have $u=\langle a \rangle $, 
 for some $a\in \mathcal{A}$. Now $u^c=\langle T_1(a)\rightarrow 0\rangle$, and hence
 $u^c\in B^{\sigma_2}$. As a result, $u\in \varrho_2$. Therefore we have $u\in \sigma_1\cap \varrho_2$. 
 Next we take $u\in \sigma_1\cap \varrho_2$, and prove that $u\in B^{\sigma_1}$. Since $B^{\sigma_1}$ is the basis for the topology $\sigma_1$, $u$ can be expressed as the union of the members of $B^{\sigma_1}$.
 As $(Hom_{\mathcal{VA}_{\mathcal{L}}}(\mathcal{A},\mathcal{L}),\sigma_1,\sigma_2)$ is pairwise compact, by Proposition \ref{PS1}, we have $u$ is compact. So $u$ can be covered by the finite collection of that members of 
 $B^{\sigma_1}$. As the finite unioin of the members of $B^{\sigma_1}$ is also in $B^{\sigma_1}$, hence $u\in B^{\sigma_1}$. Consequently, $B^{\sigma_1}=\sigma_1\cap \varrho_2$. Analogously, we can show that $B^{\sigma_2}=\sigma_2\cap \varrho_1$. Therefore, we can conclude that 
 $(Hom_{\mathcal{VA}_{\mathcal{L}}}(\mathcal{A},\mathcal{L}),\sigma_1,\sigma_2)$ is a pairwise Boolean space.

\end{proof}
\end{lem}
	We now define the functors between the categories $PBS_{\mathcal{L}}$ and $\mathcal{VA}_{\mathcal{L}}$.
\begin{defn}
\label{DS5}
We define a functor $F: PBS_{\mathcal{L}}\longrightarrow \mathcal{VA}_{\mathcal{L}}$ by 
\begin{enumerate}[(i)]
\item $F(\mathcal{B}, \alpha_{\mathcal{B}})=(Hom_{PBS_{\mathcal{L}}}((\mathcal{B},\alpha_{\mathcal{B}}),(\mathcal{L},\alpha_{\mathcal{L}})), \vee,\wedge,\rightarrow,T_p(p\in\mathcal{L}),0,1)$, where $(\mathcal{B}, \alpha_{\mathcal{B}})$ is an object of $PBS_{\mathcal{L}}$.
Here the operations $\vee,\wedge,\rightarrow,T_p(p\in\mathcal{L}),0,1$ on the set $Hom_{PBS_{\mathcal{L}}}((\mathcal{B},\alpha_{\mathcal{B}}),(\mathcal{L},\alpha_{\mathcal{L}}))$ are defined pointwise i.e., $(\phi\vee\eta)(b)=\phi(b)\vee\eta(b)$, $(\phi\wedge\eta)(b)=\phi(b)\wedge\eta(b)$, $(\phi\rightarrow\eta)(b)=\phi(b)\rightarrow\eta(b)$, $T_p(\phi)(b)=T_p(\phi(b))$, and 0,1 are considered to be constant functions.
\item $F$ acts on an arrow $\phi: (\mathcal{B},\alpha_{\mathcal{B}})\longrightarrow (\mathcal{B}',\alpha_{\mathcal{B'}})$ in $PBS_{\mathcal{L}}$ as follows:\\
 $F(\phi): F(\mathcal{B}',\alpha_{\mathcal{B'}})\longrightarrow F( \mathcal{B},\alpha_{\mathcal{B}})$ defined by $F(\phi)(\eta)=\eta\circ\phi$,\\ 
 $\eta \in Hom_{PBS_{\mathcal{L}}}((\mathcal{B}',\alpha_{\mathcal{B'}}),(\mathcal{L},\alpha_{\mathcal{L}}))$.
\end{enumerate}
\end{defn}

Note that if $\phi, \eta:((\mathcal{B},\tau_1^{\mathcal{B}},\tau_2^{\mathcal{B}}),\alpha_{\mathcal{B}})\longrightarrow ((\mathcal{L},\tau,\tau),\alpha_{\mathcal{L}})$, where $\tau$ is the discrete topology on $\mathcal{L}$, are both pairwise continuous maps, then $\phi\vee\eta$, $\phi\wedge\eta$,$\phi\rightarrow\eta$,$T_p(\phi)$ are also pairwise continuous maps. As a result, it can be shown that if $\phi,\eta\in Hom_{PBS_{\mathcal{L}}}((\mathcal{B},\alpha_{\mathcal{B}}),(\mathcal{L},\alpha_{\mathcal{L}}))$, then $\phi\vee\eta, \phi\wedge\eta, \phi\rightarrow\eta, T_p(\phi) \in Hom_{PBS_{\mathcal{L}}}((\mathcal{B},\alpha_{\mathcal{B}}),(\mathcal{L},\alpha_{\mathcal{L}}))$. Now it is being observed that $(Hom_{PBS_{\mathcal{L}}}((\mathcal{B},\alpha_{\mathcal{B}}),(\mathcal{L},\alpha_{\mathcal{L}})), \vee,\wedge,\rightarrow,T_p(p\in\mathcal{L}),0,1)$ is a $\mathcal{L}$-$\mathcal{VL}$-algebra. Now the functor $F$ is well-defined by the following Proposition \ref{PS2}.

\begin{prop}
\label{PS2}
For an arrow $\phi: (\mathcal{B},\alpha_{\mathcal{B}})\longrightarrow (\mathcal{B}',\alpha_{\mathcal{B'}})$ in $PBS_{\mathcal{L}}$, $F(\phi)$ is an arrow in $\mathcal{VA}_{\mathcal{L}}$.
\begin{proof}
	The proof is straight forward.
\end{proof}
\end{prop} 

\begin{defn}
\label{DS6}
We define a functor $G: \mathcal{VA}_{\mathcal{L}}\longrightarrow PBS_{\mathcal{L}}$ as follows:
\begin{enumerate}[(i)]
\item $G(\mathcal{A})=((Hom_{\mathcal{VA}_{\mathcal{L}}}(\mathcal{A},\mathcal{L}),\sigma_1,\sigma_2,\alpha_{\mathcal{A}})$, where $\mathcal{A}$ is an object of $\mathcal{VA}_{\mathcal{L}}$. The mapping $\alpha_{\mathcal{A}}: \Sigma_{\mathcal{L}}\longrightarrow \Omega_{Hom_{\mathcal{VA}_{\mathcal{L}}}(\mathcal{A},\mathcal{L})}$ is defined by $\alpha_{\mathcal{A}}(\mathcal{L'})=Hom_{\mathcal{VA}_{\mathcal{L}}}(\mathcal{A},\mathcal{L'})$
\item  $G$ acts on an arrow $g :\mathcal{A}_1\longrightarrow \mathcal{A}_2$ in $\mathcal{VA}_{\mathcal{L}}$ as $G(g): G(\mathcal{A}_2)\longrightarrow G(\mathcal{A}_1)$ defined by $G(g)(\mu)=\mu\circ g$, $\mu \in G(\mathcal{A}_2)$.
\end{enumerate}
\end{defn}
\begin{rem:}
In the first part of the above Definition \ref{DS6}, We take $\alpha_{\mathcal{A}}(\mathcal{L'})=Hom_{\mathcal{VA}_{\mathcal{L}}}(\mathcal{A},\mathcal{L'})$, where $\mathcal{L'}$ is a subalgebra of $\mathcal{L}$. We note that the subset $Hom_{\mathcal{VA}_{\mathcal{L}}}(\mathcal{A},\mathcal{L'})$ of $Hom_{\mathcal{VA}_{\mathcal{L}}}(\mathcal{A},\mathcal{L})$ is $\sigma_1$-closed and $\sigma_2$-closed i.e., pairwise closed, where the topologies $\sigma_1$, and $\sigma_2$ are generated by the bases $\{\langle x\rangle :x\in\mathcal{A}\}$ and $\{\langle T_1(x)\rightarrow 0 \rangle : x\in \mathcal{A}\}$, respectively. 
\end{rem:}
Now we will show the well-definedness of the functor $G$.
\begin{prop}
\label{NP3}
For an arrow $g :\mathcal{A}_1\longrightarrow \mathcal{A}_2$ in $\mathcal{VA}_{\mathcal{L}}$, $G(g): G(\mathcal{A}_2)\longrightarrow G(\mathcal{A}_1)$ is an arrow in $PBS_{\mathcal{L}}$.
\begin{proof}
For a basic open set $\langle x \rangle$, $x\in\mathcal{A}_1$ in the topology $\sigma_1^{\mathcal{A}_1}$, we get

$G(g)^{-1}(\langle x \rangle)  =  \{\phi \in Hom_{\mathcal{VA}_{\mathcal{L}}}(\mathcal{A}_2,\mathcal{L}) :G(g)(\phi)\in \langle x \rangle\} \\ 
                                        =  \{\phi \in Hom_{\mathcal{VA}_{\mathcal{L}}}(\mathcal{A}_2,\mathcal{L}) : \phi\circ g\in \langle x \rangle\} \\
                                         =  \langle g(x) \rangle \in \sigma_1^{\mathcal{A}_2}$.

Next we see that, for a basic open set $\langle x \rangle^c$, $x\in\mathcal{A}_1$, in the topology $\sigma_2^{\mathcal{A}_1}$\\ 
$G(g)^{-1}(\langle x \rangle^c)=\{\phi \in Hom_{\mathcal{VA}_{\mathcal{L}}}(\mathcal{A}_2,\mathcal{L}) : G(g)(\phi)\in \langle x \rangle^c \}\\
=\{\phi \in Hom_{\mathcal{VA}_{\mathcal{L}}}(\mathcal{A}_2,\mathcal{L}) : \phi\circ g\in \langle T_1(x)\rightarrow 0 \rangle \}\\
= \langle T_1(g(x))\rightarrow 0 \rangle \in \sigma_2^{\mathcal{A}_2}$.
  
Therefore the mapping $G(g)$ is pairwise continuous. We also observe that, if $\xi\in \alpha_{\mathcal{A}_2}(\mathcal{L'})$, then $G(g)(\xi)\in \alpha_{\mathcal{A}_1}(\mathcal{L'})$. 
As a result, $G(g)$ is an arrow in $PBS_{\mathcal{L}}$.

\end{proof}
\end{prop}
So by Lemma \ref{L1} and Proposition \ref{NP3}, we can say that $G$ is well-defined. We now establish a duality for algebras of Fitting's multi-valued logic in bitopological setting.
\begin{thm}
\label{Thm1}
The category $\mathcal{VA}_{\mathcal{L}}$ is dually equivalent to the category $PBS_{\mathcal{L}}$.
\begin{proof}
We shall prove this theorem by defining two natural isomorphisms $\beta: Id_{\mathcal{A}}\longrightarrow F\circ G$ and $\zeta: Id_{PBS_{\mathcal{L}}}\longrightarrow G\circ F$, 
where $Id_{\mathcal{A}}$ and $Id_{PBS_{\mathcal{L}}}$ are respectively the identity functors on the categories $\mathcal{VA}_{\mathcal{L}}$, and $PBS_{\mathcal{L}}$. 
Now for a $\mathcal{L}$-$\mathcal{VL}$ algebra $\mathcal{A}$, define $\beta^{\mathcal{A}}: \mathcal{A}\longrightarrow F\circ G(\mathcal{A})$ by $\beta^{\mathcal{A}}(a)(\phi)=\phi(a)$, 
$a\in\mathcal{A}$ and $\phi\in G(\mathcal{A})=Hom_{\mathcal{VA}_{\mathcal{L}}}(\mathcal{A},\mathcal{L})$. It is easily seen that $\beta^{\mathcal{A}}$ is a homomorphism. Using Theorem \ref{}
 and Lemma \ref{}, it can be shown that $\beta^{\mathcal{A}}$ is one-one and onto. Consequently, $\beta^{\mathcal{A}}$ is an isomorphism. \\
Again for an object $(\mathcal{S},\alpha_{\mathcal{S}})$ of $PBS_{\mathcal{L}}$, define $\zeta_{(\mathcal{S},\alpha_{\mathcal{S}})}: (\mathcal{S},\alpha_{\mathcal{S}})\longrightarrow G\circ F
(\mathcal{S},\alpha_{\mathcal{S}})$ by $\zeta_{(\mathcal{S},\alpha_{\mathcal{S}})}(s)(\psi)=\psi(s)$, where $s\in\mathcal{S}$ and 
$\psi\in F((\mathcal{S},\alpha_{\mathcal{S}}))=(Hom_{PBS_{\mathcal{L}}}((\mathcal{S},\alpha_{\mathcal{S}}),(\mathcal{L},\alpha_{\mathcal{L}})), \vee,\wedge, \rightarrow,T_p(p\in\mathcal{L}),0,1)$. 
We shall show that $\zeta_{(\mathcal{S},\alpha_{\mathcal{S}})}$ is a bi-homeomorphism.
As $\psi \in Hom_{PBS_{\mathcal{L}}}((\mathcal{S},\alpha_{\mathcal{S}}),(\mathcal{L},\alpha_{\mathcal{L}}))$, so for each $s\in \mathcal{S}$, $\zeta_{(\mathcal{S},\alpha_{\mathcal{S}})}(s)$ 
is a $\mathcal{L}$-$\mathcal{VL}$-algebra homomorphism. Henceforth $\zeta_{(\mathcal{S},\alpha_{\mathcal{S}})}$ is well-defined. To prove the pairwise continuity of 
$\zeta_{(\mathcal{S},\alpha_{\mathcal{S}})}$, we show that $\zeta_{(\mathcal{S},\alpha_{\mathcal{S}})}^{-1}({\langle v \rangle}), 
v\in Hom_{PBS_{\mathcal{L}}}((\mathcal{S},\alpha_{\mathcal{S}}),(\mathcal{L},\alpha_{\mathcal{L}}))$ is $\tau_1$-open and $\zeta_{(\mathcal{S},\alpha_{\mathcal{S}})}^{-1}{(\langle v \rangle^c)}, 
v\in Hom_{PBS_{\mathcal{L}}}((\mathcal{S},\alpha_{\mathcal{S}}),(\mathcal{L},\alpha_{\mathcal{L}}))$ is $\tau_2$-open. Now $\zeta_{(\mathcal{S},\alpha_{\mathcal{S}})}^{-1}({\langle v \rangle}) 
=\{s\in\mathcal{S}: \zeta_{(\mathcal{S},\alpha_{\mathcal{S}})}(s)\in \langle v \rangle \}=\{s\in\mathcal{S}:\zeta_{(\mathcal{S},\alpha_{\mathcal{S}})}(s)(v)=1\}=v^{-1}\{1\}$. As $v^{-1}\{1\}$ is 
$\tau_1$-open, so $\zeta_{(\mathcal{S},\alpha_{\mathcal{S}})}^{-1}({\langle v \rangle})$ is open.
$\zeta_{(\mathcal{S},\alpha_{\mathcal{S}})}^{-1}{(\langle v \rangle^c)}=\{s\in\mathcal{S}: \zeta_{(\mathcal{S},\alpha_{\mathcal{S}})}(s)\in \langle T_1(v)\rightarrow 0 \rangle \}
=\{s\in\mathcal{S}: \zeta_{(\mathcal{S},\alpha_{\mathcal{S}})}(s)(T_1(v)\rightarrow 0)=1\}=(T_1(v)\rightarrow 0)^{-1}\{1\}$. Since $(T_1(v)\rightarrow 0)^{-1}\{1\}$ is $\tau_2$-open, henceforth 
$\zeta_{(\mathcal{S},\alpha_{\mathcal{S}})}^{-1}{(\langle v \rangle^c)}$ is $\tau_2$-open.\\
For any two points $s, s'\in\mathcal{S}$, assume $\zeta_{(\mathcal{S},\alpha_{\mathcal{S}})}(s)=\zeta_{(\mathcal{S},\alpha_{\mathcal{S}})}(s')$. 
Since $\{\langle v\rangle: v\in Hom_{PBS_{\mathcal{L}}}((\mathcal{S},\alpha_{\mathcal{S}}),(\mathcal{L},\alpha_{\mathcal{L}}))\}$, and $\{\langle v \rangle^c:v\in Hom_{PBS_{\mathcal{L}}}((\mathcal{S},\alpha_{\mathcal{S}}),(\mathcal{L},\alpha_{\mathcal{L}}))\}$ are the bases for the topologies $\sigma_1$ and $\sigma_2$, respectively on $G\circ F(\mathcal{S},\alpha_{\mathcal{S}})$. So for some 
$v\in Hom_{PBS_{\mathcal{L}}}((\mathcal{S},\alpha_{\mathcal{S}}),(\mathcal{L},\alpha_{\mathcal{L}}))$, either $\zeta_{(\mathcal{S},\alpha_{\mathcal{S}})}(s)\in \langle v \rangle$ or 
$\zeta_{(\mathcal{S},\alpha_{\mathcal{S}})}(s)\in \langle v \rangle^c$. In both the cases we get $v(s)=v(s')$. Since $\mathcal{S}$ is pairwise Hausdorff, we get $s=s'$. Consequently,
$\zeta_{(\mathcal{S},\alpha_{\mathcal{S}})}$ is one-one. We now show that $\zeta_{(\mathcal{S},\alpha_{\mathcal{S}})}$ is onto. Let $\phi\in G\circ F(\mathcal{S},\alpha_{\mathcal{S}})$.
Define $S_1=\{v^{-1}(\{1\})\in \beta_1: \phi(v)=1 \}$ and $S_2=\{v^{-1}(\{0\}) \in\beta_2: \phi(v)=0\}$, where $\beta_1$, $\beta_2$ are bases for the topologies $\tau_1$, and $\tau_2$, respectively.
We show that $S_1\cup S_2$ has the finite intersetion property. Since $(S_1\cup S_2)\cap (S_1'\cup S_2')=(S_1\cap S_1')\cup (S_2\cap S_2')$, and $v^{-1}(\{1\})\cap v'^{-1}(\{1\})
=(v\wedge v')^{-1}(\{1\})$, $v^{-1}(\{0\})\cap v'^{-1}(\{0\})=(v\wedge v')^{-1}(\{0\})$, we only show that $v^{-1}(\{1\})\neq \emptyset$ and $v^{-1}(\{0\})\neq\emptyset$ under the given conditions.
Suppose $\phi(v)=1$ but $v^{-1}(\{1\})=\emptyset$. Then we have $T_1(v)=0$, and henceforth $T_1(\phi(v))=\phi(T_1(v))=0\Rightarrow \phi(v)=0$. Again take $\phi(v)=0$ but $v^{-1}(\{0\})=\emptyset$, then we have $T_0(v)=0$. Therefore $\phi(T_0(v))=T_0(\phi(v))=0\Rightarrow \phi(v)\neq 0$, but it contradicts the given assumption. As $\mathcal{S}$ is pairwise compact and pairwise Hausdorff,
there exists $s\in\mathcal{S}$ such that $\{s\}=\bigcap(S_1\cup S_2)$. Thus we have, $\zeta_{(\mathcal{S},\alpha_{\mathcal{S}})}(s)=\phi$. 
Therefore $\zeta_{(\mathcal{S},\alpha_{\mathcal{S}})}$ is onto.\\
Finally we show that $\zeta_{(\mathcal{S},\alpha_{\mathcal{S}})}^{-1}$ is pairwise continuous. It can only be shown that $\zeta_{(\mathcal{S},\alpha_{\mathcal{S}})}$ is bi-closed. Let $U$ be a $\tau_1$-closed set. Since $(\mathcal{S},\tau_1,\tau_2)$ is a pairwise Boolean space, both $(\mathcal{S},\tau_1)$ and $(\mathcal{S},\tau_2)$ are compact. Consequently $\zeta_{(\mathcal{S},\alpha_{\mathcal{S}})}(U)$ is compact in
$(Hom_{{\mathcal{VA}}_{\mathcal{L}}}(Hom_{PBS_{\mathcal{L}}}((\mathcal{S},\alpha_{\mathcal{S}}),(\mathcal{L},\alpha_{\mathcal{L}})),\mathcal{L})$. We observe that the topological space $(Hom_{{\mathcal{VA}}_{\mathcal{L}}}(Hom_{PBS_{\mathcal{L}}}((\mathcal{S},\alpha_{\mathcal{S}}),(\mathcal{L},\alpha_{\mathcal{L}})),\mathcal{L}),\sigma_1)$ with basis $\{\langle v \rangle :v\in  Hom_{PBS_{\mathcal{L}}}((\mathcal{S},\alpha_{\mathcal{S}}),(\mathcal{L},\alpha_{\mathcal{L}}))\}$  is itself a Hausdorff space, and thus $\zeta_{(\mathcal{S},\alpha_{\mathcal{S}})}(U)$ is closed. The topological space $(Hom_{{\mathcal{VA}}_{\mathcal{L}}}(Hom_{PBS_{\mathcal{L}}}((\mathcal{S},\alpha_{\mathcal{S}}),(\mathcal{L},\alpha_{\mathcal{L}})),\mathcal{L}),\sigma_2)$ with basis $\{\langle T_1(v)\rightarrow 0 \rangle :v\in  Hom_{PBS_{\mathcal{L}}}((\mathcal{S},\alpha_{\mathcal{S}}),(\mathcal{L},\alpha_{\mathcal{L}}))\}$ is itself a Hausdorff space, so for a $\tau_2$-closed set $U'$, $\zeta_{(\mathcal{S},\alpha_{\mathcal{S}})}(U')$ is closed. Therefore $\zeta_{(\mathcal{S},\alpha_{\mathcal{S}})}$ is a bi-homeomorphism.

\end{proof}
\end{thm}

\section{Duality for $\mathbb{I}\mathbb{S}\mathbb{P_I}(\mathcal{L})$:}\label{sec:3}
 \subsection{The concept of $\mathbb{I}\mathbb{S}\mathbb{P_I}(\mathcal{L})$:}
 Here $\mathcal{L}$ denotes a finite algebra equipped with join and meet operations, and top element 1, bottom element 0, in which underlying lattice is finite bounded distributive . We understand a intuitionistic Kripke frame by a tuple $(W,R)$, where $W$ is a non-empty set and $R$ is a partial order relation on $W$.
 For any two elemnts $r_1,r_2\in \mathcal{L}$, $r_1\rightarrow r_2=\bigvee\{\ell\in\mathcal{L}:r_1\wedge\ell\leq r_2\}$. $\mathbb{ISP}(\mathcal{L})$ symbolized as the class of all isomorphic copies of subalgebras of direct powers of finite single algebra $\mathcal{L}$.
 It follows from \cite{clark1998natural}, that for the 2-element distributive lattice $\{0,1\}$, $\mathbb{ISP}(\{0,1\})$ coincides with the class of distributive lattices.\\
 
 The category $ISP(\mathcal{L})$ is defined as follows.
 \begin{defn}
 	\label{Df1}
 
 Objects of the category are taken as an algebras in $\mathbb{ISP}(\mathcal{L})$ and arrows are the homomorphisms between the objects, which respect the operations defined on $\mathcal{L}$.
 \end{defn}

\begin{defn}[\cite{maruyama2012natural}]
	\label{D0}
The intuitionistic power of $\mathcal{L}$ with respect to a intutionistic frame $(W,R)$ is defined as $\mathcal{L}^W \in \mathbb{I}\mathbb{S}\mathbb{P}(\mathcal{L})$ equipped with the binary operation $\rightarrow$
(intuionistic implication) on $\mathcal{L}^W$ defined as $(f\rightarrow g)(w)=\bigwedge\{f(w')\rightarrow g(w'): wRw'\}$, where $f, g\in \mathcal{L}^W$.
\end{defn}                                                                                                                                                                                                       	                                                                                                                                                                                                                                                                                                                                                                                                                                                                                                                                                                                                                                                                                                                                                                                                                                                                                                                                                                                                                                                                                                                                                                                                                                                                                                                                                                                                                                                                                                                                                                                                                                                                                                                                                                           
                                                               
The concept of $\mathbb{I}\mathbb{S}\mathbb{P_I}(\mathcal{L})$ is given in the following definition.
\begin{defn}
$\mathbb{I}\mathbb{S}\mathbb{P_I}(\mathcal{L})$ is denoted by the class of all isomorphic copies of subalgebras of intuitionistic power of $\mathcal{L}$.
\end{defn}

The category $ISP_I(\mathcal{L})$ is defined as follows:
\begin{defn}
\label{D}
Objects of the category $ISP_I(\mathcal{L})$ are the algebras in $\mathbb{I}\mathbb{S}\mathbb{P_I}(\mathcal{L})$ and arrows are homomorphisms which are defined by a function  between the objects of $ISP_I(\mathcal{L})$ such that preserve the implication operation $\rightarrow$ and 
all the other operations of $\mathcal{L}$.
\end{defn}
\begin{defn}
	\label{D1}
	We define an order relation $R$ on $Hom_{ISP(\mathcal{L})}(\mathcal{A},\mathcal{L})$ as for any $v_1, v_2\in Hom_{ISP(\mathcal{L})}(\mathcal{A},\mathcal{L})$, $v_1Rv_2$ iff $v_1(x)\leq v_2(x)$, for all $x\in \mathcal{A}$. Then $(Hom_{ISP(\mathcal{L})}(\mathcal{A},\mathcal{L}),R)$ is a poset.
\end{defn}
 
\begin{defn}
	\label{D2}
	For any object $(\mathcal{A},\rightarrow)$ of $ISP_I(\mathcal{L})$, and $v\in Hom_{ISP(\mathcal{L})}(\mathcal{A},\mathcal{L})$, $\mathbb{I}\mathbb{S}\mathbb{P_I}(\mathcal{L})$ satisfies the intuitionistic Kripke model condition iff $v(x\rightarrow y)=\bigwedge\{w(x)\rightarrow_{\mathcal{L}}w(y):vRw\}$
\end{defn}
Now it can be easily observe the following Proposition \ref{D3}.
\begin{prop}[\cite{clark1998natural}]
	\label{D3}
	For 2-element distributive lattice $\{0,1\}$, $\mathbb{ISP_{I}}(\{0,1\})$ coincides with the class of all Heyting algebras.
\end{prop}
A zero-dimensional compact Hausdorff space is called a Stone space. An ordered topological-space is defined as a triple $(X,\tau,R)$, where the tuple $(X,\tau)$ is a topological-space and $(X,R)$ is a partially ordered set. For an ordered set $(X,R)$ we have $R(x)=\{y\in X: xRy\}$ and $R^{-1}(X_0)=\{y\in X: yRx, \text{for some x}\in X_0\}$, where $X_0\subset X$. Then $R(x)$ is an up-set, and $R^{-1}(X_0)$ is a down-set.

\subsection{Duality for $\mathbb{ISP}({\mathcal{L}})$}
\begin{defn}
	\label{P1}
	Let $A\in ISP(\mathcal{L})$, and $P$ be a non-void subset of $A$. $P$ is said to be a filter, if $P$ is an up-set and for any $a,b\in P$, $a\wedge b\in P$. Dual of this notion gives ideal.
\end{defn}

\begin{defn}
	\label{P2}
	Let $A\in ISP(\mathcal{L})$, and $P$ be a filter of $A$. $P$ is called a prime filter if $P\neq A$ and for any element $a,b$ of $A$, $a\vee b\in P\Rightarrow$ either $a\in P$ or $b\in P$. 
	
\end{defn}
\begin{prop}
	\label{P3}
	Let $A\in \mathbb{ISP}(\mathcal{L})$, and $\mathcal{P}$ be a prime-ideal of $\mathcal{L}$. Then there is a one-one correspondence $f: PI_{\mathcal{L}}\longrightarrow HOM_{ISP(\mathcal{L})}(A)$ such that $f(\mathcal{P})=V_{\mathcal{P}}$, with $V_{\mathcal{P}}^{-1}(0)=\mathcal{P}$. Define for each $x\in A$, $V_{\mathcal{P}}(x)=r$ iff $\chi_r(x)\notin \mathcal{P}$. $\chi$ is a characteristic function, and $PI_{\mathcal{L}}$ denotes the set of prime-ideals of $\mathcal{L}$.
\end{prop}
We now recall a result from \cite{stone1937topological}

\begin{thm}[\cite{stone1937topological}]
	\label{P4}
	Let $A$ be a distributive lattice and $x,y$ be any members of $A$. If $x\neq y$, then there is a prime ideal $\mathcal{P}$ such that $x\in\mathcal{P}$ and $y\notin\mathcal{P}$.
\end{thm}



\begin{defn}
\label{D4}
We take a category $PSpa$ as follows:
\begin{enumerate}
	\item Object: An object of $PSpa$ is taken as a triple $(X,\tau,R)$, where $(X,\tau)$ is a compact space, and $R$ is a partial order relation on $X$ such that the following condition hold:
\begin{enumerate}[(i)]
	\item if $x\not R y$, then for some clopen up-set $W$ of $X$ such that $x\in W$ but $y\notin W$
	
\end{enumerate}
\item Arrow: An arrow $\psi : (X,\tau_1,R_1)\longrightarrow (Y,\tau_2,R_2)$ in $PSpa$ is taken as a continuous map $\psi :(X,\tau_1)\longrightarrow (Y,\tau_2)$, and which is order preserving i.e., for any $x,y\in X$, if $xR_1y$ then $\psi(x)R_2\psi(y)$.

\end{enumerate}

\end{defn}
Consider $\mathcal{L}$ is taken with the discrete topology, and $\leq$ is partial order relation on $\mathcal{L}$ defined by $\ell_1\leq \ell_2$ if $\ell_1 \vee \ell_2=\ell_2$, and $\ell_1\wedge\ell_2=\ell_1$.

For $A\in \mathbb{ISP}(\mathcal{L})$, taking a ordered topological space $(HOM_{ISP(\mathcal{L})}(A,\mathcal{L}),\tau,R)$ where the topology $\tau$ is generated by $\{<a>: a\in A\}$, $<a>=\{v\in HOM_{ISP(\mathcal{L})}(A,\mathcal{L}): v(a)=1\}$. Now for each $a\in A$, the set $<a>$ is a clopen up-set. 
\begin{defn}
	\label{D5}
We define a functor $\mathcal{G}:ISP(\mathcal{L})\longrightarrow PSpa$ as follows:
\begin{itemize}
\item $\mathcal{G}$ acts on an object $A$ of $ISP(\mathcal{L})$ as $\mathcal{G}(A)=(HOM_{ISP(\mathcal{L})}(A,\mathcal{L}),R_A)$,
	\item $\mathcal{G}$ acts on an arrow $f:A\longrightarrow B$ in $ISP(\mathcal{L})$ as $\mathcal{G}(f):\mathcal{G}(B)\longrightarrow \mathcal{G}(A)$ defined by $\mathcal{G}(f)(\phi)=\phi\circ f$, $\phi\in\mathcal{G}(B)$.
\end{itemize}

\end{defn}

We shall now show the well-definedness of the functor $\mathcal{G}$.
\begin{lem}
	\label{D6}
	For each object $A$ of $ISP(\mathcal{L})$, $(HOM_{ISP(\mathcal{L})}(A,\mathcal{L}),R_A)$ is an object of $PSpa$.
	\begin{proof}
		The set $HOM_{ISP(\mathcal{L})}(A,\mathcal{L})$ is compact as $\mathcal{L}^A$ with product topology is compact, and $HOM_{ISP(\mathcal{L})}(A,\mathcal{L})$ is closed in the defined topology $\tau$,  which can be induced by the product topology on $\mathcal{L}^A$.\\
		Now we see that if $v\not R_A w$, then there exists $a\in A$ such that $v(a)=1$ and $w(a)\neq 1$. Then $v\in <a>$ and $w\in <a>^c$.
		
	\end{proof}

\end{lem}

\begin{lem}
	\label{D7}
	For an arrow $f$ in $ISP(\mathcal{L})$, $\mathcal{G}(f)$ is an arrow in $PSpa$.
	\begin{proof}
	Here an arrow $f:A\longrightarrow B$ in $ISP(\mathcal{L})$. $\mathcal{G}$ acts as $\mathcal{G}(f):\mathcal{G}(B)\longrightarrow \mathcal{G}(A)$ defined by $\mathcal{G}(f)(\phi)=\phi\circ f$. It is observed that, for each $a\in A$, $\mathcal{G}(f)^{-1}(<a>)=\{v\in \mathcal{G}(B): v\circ f(a)=1\}=<f(a)>$. Henceforth, $\mathcal{G}(f)$ is continuous.\\
	Now for any two members $v,w$ of $\mathcal{G}(B)$ if $vR_Bw$, then $v(b)\leq w(b)$, $\forall b\in B$. We get $v(f(a))\leq w(f(a))$, and thus $\mathcal{G}(f)(v)R_A\mathcal{G}(f)(w)$.
	\end{proof}
\end{lem}

Therefore the functor $\mathcal{G}$ is well-defined by Lemma \ref{D6}, and Lemma \ref{D7}.
\begin{defn}	\label{D8}
	We define a functor $\mathcal{C}: PSpa\longrightarrow ISP(\mathcal{L})$ as follows:
	\begin{itemize}
		\item $\mathcal{C}$ acts on an object $(S,R)$ of $PSpa$ as $\mathcal{C}(S,R)=HOM_{PSpa}((S,R),(\mathcal{L},\leq))$.
		\item $\mathcal{C}$ acts on an arrow $f:(S_1,R_1)\longrightarrow (S_2,R_2)$ in $PSpa$ as $\mathcal{C}(f):\mathcal{C}(S_2,R_2)\longrightarrow \mathcal{C}(S_1,R_1)$ defined by $\mathcal{C}(f)(\phi)=\phi\circ f$, $\phi\in \mathcal{C}(S_2,R_2)$. 
	\end{itemize}

\end{defn}

\begin{note}
	\label{N1}
	For each object $(S,R)$ of $PSpa$, the set $HOM_{PSpa}((S,R),(\mathcal{L},\leq))$ is taken with the operations $(\vee,\wedge,0,1)$ and are defined pointwise i.e., for any $f,g\in HOM_{PSpa}((S,R),(\mathcal{L},\leq))$, $(f\vee g)(s)=f(s)\vee g(s)$, $(f\wedge g)(s)=f(s)\wedge g(s)$. Operations 0,1 are taken as constant functions.
\end{note}

Well-definedness of the functor $\mathcal{C}$ is shown by the following Lemma \ref{D9}.
\begin{lem}
	\label{D9}
	For an arrow $f:(S_1,R_1)\longrightarrow (S_2,R_2)$ in $PSpa$, $\mathcal{C}(f)$ is an arrow in $ISP(\mathcal{L})$.
	\begin{proof}
		Well-definedness of the map $\mathcal{C}(f)$ is followed by the definition and as $\phi$ is an order preserving continuous map. It also preserves all the defined operations.
	\end{proof}
\end{lem}

\begin{thm}
	\label{D10}
	For an object $A$ of $ISP(\mathcal{L})$, $A$ is isomorphic to $\mathcal{C}\circ \mathcal{G}(A)$ in $ISP(\mathcal{L})$
	\begin{proof}
	Define a map $\sigma_A: A\longrightarrow \mathcal{C}\circ \mathcal{G}(A)$ by $\sigma_A(a)(v)=v(a)$, where $a\in A$, and $v\in \mathcal{G}(A)$. Now it is easily observed that, for each $a\in A$, $\sigma_A(a)\in \mathcal{C}\circ \mathcal{G}(A)$.For each $a\in A$, and $r(\neq 1)\in\mathcal{L}$, $\sigma_A(a)^{-1}\{r\}=\{v\in HOM_{ISP(\mathcal{L})}(A,\mathcal{L}):v(a)=r\}\subset \langle a \rangle^c$. Similarly for $a\in A$, $\sigma_A(a)^{-1}\{1\}=\langle a \rangle$. Henceforth, $\{\sigma_A(a):a\in A\}\subset \mathcal{C}\circ\mathcal{G}(A)$

	 So $\sigma_A$ is well-defined. $\sigma_A$ is a homomorphism as the operations are defined point-wise on $\mathcal{C}\circ \mathcal{G}(A)$.\\
	We show that $\sigma_A$ is one-one. For any members $a,b\in A$, if $a\neq b$, then we claim $\sigma_A(a)\neq\sigma_A(b)$. Then it is required to show that $v(a)\neq v(b)$. It can be shown by Proposition \ref{P3} and Theorem \ref{P4}.\\
	Now if $\psi\in\mathcal{C}\circ\mathcal{G}(A)$, we claim that $\psi=\sigma_A(a)$, for some $a\in A$. Now $\psi^{-1}(\{1\})$ is a clopen up-set of $\mathcal{G}(A)$, and then $\psi^{-1}(\{1\})=\bigvee_{i\in I}\langle a_i\rangle$. So $HOM_{ISP(\mathcal{L})}(A,\mathcal{L})=\bigvee_{i\in I}\langle a_i\rangle \bigvee (\psi^{-1}(1))^c$. As $HOM_{ISP(\mathcal{L})}(A,\mathcal{L})$ is compact, so we may consider $HOM_{ISP(\mathcal{L})}(A,\mathcal{L})=\bigvee^{n}_{i=1} \langle a_i \rangle \bigvee (\psi^{-1}(1))^c$. Therefore, $\psi^{-1}(\{1\})\leq \bigvee^{n}_{i=1} \langle a_i \rangle$, and hence $\psi^{-1}(\{1\})=\bigvee_{i\in J}\langle a_i\rangle=\langle \bigvee_{i\in J}a_i\rangle$. Let $a=\bigvee^{n}_{i=1}a_i$, and if $v\in\psi^{-1}(\{1\})$ then $\sigma_A(a)(v)=\psi(v)$. Therefore  $\psi=\sigma_A(a)$, and hence $\sigma_A$ is surjective.

	\end{proof}
\end{thm}

\begin{thm}
	\label{D11}
	For an object $(S,R)$ in $PSpa$, $(S,R)$ is homeomorphic to $\mathcal{G}\circ\mathcal{C}(S,R)$.
	\begin{proof}
	Define a map $\delta_S: (S,R)\longrightarrow \mathcal{G}\circ\mathcal{C}(S,R)$ by $\delta_S(s)(f)=f(s)$. For each $s\in S$, $\delta_S(s)$ is a homomorphism, as the operations are defined pointwise on $\mathcal{C}(S,R)$. Therefore, $\delta_S$ is well-defined.\\
	Now we observe that if $f\in\mathcal{C}(S,R)$, then $\delta_S^{-1}(\langle f\rangle)=\{s\in S:\delta_S(s)(f)=1\}=f^{-1}\{1\}$, is an open up-set in $(S,R)$. $\delta_S$ is also order preserving map, since if $s_1Rs_2$ then $f(s_1)\leq f(s_2)$, and henceforth $\delta_S(s_1)R'\delta_S(s_2)$, where $R'$ is taken as a  partially order relation accroding to Definition \ref{D1} on $\mathcal{G}\circ\mathcal{C}(S,R)$.\\
	Let $s\neq t$ in $S$. We claim that $\delta_S(s)\neq\delta_S(t)$. The claim is shown by the fact that $\mathcal{G}\circ\mathcal{C}(S,R)$ is an object of $PSpa$ and hence it is Hausdorff and zero-dimensional. Thus there exists $\phi\in\mathcal{C}(S,R)$ such that $\phi(s)\neq\phi(t)$, hence $\delta_S$ is one-one.
	
    We now show that $\delta_S$ is surjective.
	We already observe that $\{\delta_S(s):s\in S\}\subset \mathcal{G}\circ\mathcal{C}(S,R)$. As $\delta_S(S)$ is a compact subset of $\mathcal{G}\circ\mathcal{C}(S,R)$ and hence $\delta_S(S)$ is 
closed. 
If $\delta_S$ is not surjective, then there exists $v\in \mathcal{G}\circ\mathcal{C}(S,R)$ such that $v\notin \delta_S(S)$ i.e., $v\neq\delta_S(s)$, for any $s\in S$. 
Therefore there exists a clopen up-set $W$ in $\mathcal{G}\circ\mathcal{C}(S,R)$ containing $v$ but missing $\delta_S(S)$. As $W$ is compact, so $W$ can be expressed as finite union of basis open sets. We may consider $W=\langle f\rangle\wedge \langle g\rangle^c$, 
for some $f,g\in \mathcal{C}(S,R)$. Now $\delta_S^{-1}(W)=\delta_S^{-1}(\langle f\rangle)\wedge \delta_S^{-1}(\langle g\rangle^c)$. Since $\delta_S^{-1}(W)=\emptyset$, 
hence $\delta_S^{-1}(\langle f \rangle)\subset \delta_S^{-1}(\langle g \rangle)$. If $\langle f\rangle\subset \langle g \rangle$, then $W=\emptyset$, otherwise for some $s\in S$, we get $\delta_S(s)\in W$.\\
It is easy to observe that $\delta_S$ is a closed map. Now we show that for any members $s_1,s_2\in S$, if $\delta_S(s_1) R'\delta_S(s_2)$ then $s_1Rs_2$. We show the equivalent statement i.e., if $s_1\not R s_2$, then $\delta_S(s_1) \not R'\delta_S(s_2)$. Since $s_1\not R s_2$, then there exists a clopen set $\mathcal{U}\subset S$ such that $s_2\in\mathcal{U}$ and $\mathcal{U}\cap R(s_1)=\emptyset$. Define $f:S\longrightarrow \mathcal{L}$ by \[f(s)= \left\{
\begin{array}{ll}
0 & s\in\mathcal{U} \\
1 & s\notin\mathcal{U} \\
\end{array} 
\right. \]
Then $f(s_1)\nleq f(s_2)$, and hence $\delta_S(s_1)(f)\not R'\delta_S(s_2)(f)$. Therefore for any $s_1, s_2\in S$, $s_1Rs_2$ iff $\delta_S(s_1)R'\delta_S(s_2)$. Now $\delta_S^{-1}$ is order preserving, since $\delta_S$ is bijective, and the relation $s_1Rs_2\iff\delta_S(s_1)R'\delta_S(s_2)$ hold.

\end{proof}
\end{thm}

\begin{thm}
	\label{D12}
	The category $ISP(\mathcal{L})$ is dually equivalent to the category $PSpa$.
	\begin{proof}
	Using Theorem \ref{D10} and Theorem \ref{D11}, we can conclude that the category $ISP(\mathcal{L})$ is dually equivalent to the category $PSpa$.
	\end{proof} 
	
\end{thm}
Now we develope a duality for $\mathbb{ISP_I}(\mathcal{L})$ using Theorem \ref{D12}.
\begin{defn}
	\label{D13}
	We take a category $HSpa$ as follows.
	\begin{enumerate}
		\item Object: An object of $HSpa$ is taken as a triple $(S,\tau,R)$ such that $(S,\tau,R)$ is an object of $PSpa$ which additionaly satisfies the following condition:
		\begin{enumerate}[(i)]
			\item if $\mathcal{C}$ is a clopen subset of $S$, then $R^{-1}(\mathcal{C})$ is a clopen down set of $S$.
			\end{enumerate}
		\item Arrow: An arrow $\phi:(S_1,\tau_1,R_1)\longrightarrow (S_2,\tau_2,R_2)$ in $HSpa$ is an arrow in $PSpa$ which satisfies the following condition:
		\begin{enumerate} [(i)]
			\item for any members $s_1\in S_1$, and $s_2\in S_2$, if $\phi(s_1)R_2s_2$ then there exists $s\in S_1$ such that $s_1R_1s$ and $\phi(s)=s_2$.
			\end{enumerate}
		
	\end{enumerate}
\end{defn}

\begin{defn}
	\label{D14}
	We define a functor $\mathcal{G}_I:ISP_I(\mathcal{L})\longrightarrow HSpa$ as follows:
	\begin{itemize}
		\item $\mathcal{G}_I$ acts on an object $(A,\rightarrow)$ of $ISP_I(\mathcal{L})$ as $\mathcal{G}_I(A)=(HOM_{ISP_I(\mathcal{L})}(A,\mathcal{L}),R_A)$,
		\item $\mathcal{G}_I$ acts on an arrow $f:A\longrightarrow B$ in $ISP_I(\mathcal{L})$ as $\mathcal{G}_I(f):\mathcal{G}_I(B)\longrightarrow \mathcal{G}_I(A)$ defined by $\mathcal{G}_I(f)(\phi)=\phi\circ f$, $\phi\in\mathcal{G}_I(B)$.
	\end{itemize}
\end{defn}
\begin{lem}
	\label{D15}
	For an object $(A,\rightarrow)$ of $ISP_I(\mathcal{L})$, $\mathcal{G}_I(A)$ is an object of $HSpa$.
	\begin{proof}
	By Lemma \ref{D6}, $\mathcal{G}_I(A)$ is an object of $PSpa$. We show that for each clopen subset $\mathcal{U}$ of $S$, $R_A^{-1}(\mathcal{U})$ is also clopen. Since $\{\langle a \rangle:a\in A\}$ is a clopen basis for the topology on $HOM_{ISP(\mathcal{L})}(A,\mathcal{L})$, and $R_A^{-1}$ preserves union , therefore we show that $R_A^{-1}(\langle a \rangle)$ is clopen, for each $a\in A$. Now we observe that $R_A^{-1}(\langle a \rangle)=\langle a\rightarrow 0 \rangle^c$. If $v\in \langle a\rightarrow 0 \rangle^c$, then $v(a\rightarrow 0)\neq 1$. Now $v(a\rightarrow 0)=\bigwedge\{u(a\rightarrow 0):vR_Au\}=\bigwedge\{u(a)\rightarrow 0:vR_Au\}$. Since $u(a)\rightarrow 0=0 \text{or} 1$, hence there exists $u\in HOM_{ISP(\mathcal{L})}(A,\mathcal{L})$ such that $vR_Au$ and $u(a)\rightarrow 0=0$. Thus $u(a)=1$, henceforth $v\in R_A^{-1}(\langle a \rangle)$. Again if $v\in R_A^{-1}(\langle a \rangle)$, then there exists $u\in \langle a \rangle$ such that $vR_A u$. So $v(a\rightarrow 0)=0$. Hence $v\in \langle a \rangle^c$.

	\end{proof}
	
\end{lem}
\begin{lem}
	\label{D16}
	For an arrow $f:(A,\rightarrow)\longrightarrow (B,\rightarrow)$ in $ISP_I(\mathcal{L})$, $\mathcal{G}_I(f)$ is an arrow in $HSpa$.
	\begin{proof}
	Here $\mathcal{G}_I(f):\mathcal{G}_I(B)\longrightarrow \mathcal{G}_I(A)$ is defined as $\mathcal{G}_I(f)(\phi)=\phi\circ f$, where $\phi\in\mathcal{G}_I(B)$. Then by Lemma \ref{D9}, $\mathcal{G}_I(f)$ is an arrow in $PSpa$. Now we show the condition given in the arrow part of Definition \ref{D13}. We show the equivalent condition that $\mathcal{G}_I(f)(R_B(v))=R_A(\mathcal{G}_I(f)(v))$, $\forall v\in\mathcal{G}_I(B)$. We observe that if $\psi\notin\mathcal{G}_I(f)(R_B(v))$ then $\psi\notin R_A(\mathcal{G}_I(f)(v))$. Since $\psi\notin\mathcal{G}_I(f)(R_B(v))$, then $\psi\neq\mathcal{G}_I(f)(w)$, for any $w\in HOM_{ISP(\mathcal{L})}(B,\mathcal{L})$ such that $vR_Bw$.	Then by the object part of  Definition \ref{D4}, we can take $\psi\in \langle a \rangle$ and  $\mathcal{G}_I(f)(w)\in \langle a\rangle^c$. Now if $\psi\in R_A(\mathcal{G}_I(f)(v)))$, then $(v\circ f)R_A\psi$. Therefore by definition of $R_A$, $(v\circ f)(a)\leq \psi(a)$, $\forall a\in A$. Then $(v\circ f)(a\rightarrow 0)\leq \psi(a\rightarrow 0)$. But $\psi(a\rightarrow 0)=\psi(a)\rightarrow 0=0$, as $\psi(a)=1$. Now $\mathcal{G}_I(f)(v)(a\rightarrow 0)=\bigwedge\{\mathcal{G}_I(f)(w)(a\rightarrow 0):vR_Bw\}$. Since $\mathcal{G}_I(f)(v)(a\rightarrow 0)=0$, hence there exists $w\in HOM_{ISP(\mathcal{L})}(B,\mathcal{L})$ such that $vR_Bw$ and $\mathcal{G}_I(f)(w)(a\rightarrow 0)=0$. We see that $\mathcal{G}_I(f)(w)(a\rightarrow 0)=\mathcal{G}_I(f)(w)(a)\rightarrow 0=0\Rightarrow \mathcal{G}_I(f)(w)(a)=1$.  This contradicts the assumption that $\mathcal{G}_I(f)(w)\in \langle a\rangle^c$. Hence $\psi\notin R_A(\mathcal{G}_I(f)(v))$.
	Thus equivalently we have, if $\psi \in R_A(\mathcal{G}_I(f)(v))$ then $\psi\in \mathcal{G}_I(f)(R_B(v))$. Therefore $R_A(\mathcal{G}_I(f)(v))\subseteq \mathcal{G}_I(f)(R_B(v))$.
	It is easy to observe that $\mathcal{G}_I(f)(R_B(v))\subseteq R_A(\mathcal{G}_I(f)(v))$.

	\end{proof}
\end{lem}

Therefore the functor $\mathcal{G}_I$ is well-defined by Lemma \ref{D15}, and Lemma \ref{D16}.
\begin{defn}
	\label{D17}
	We define a functor $\mathcal{C}_I:HSpa\longrightarrow ISP_I(\mathcal{L})$ as follows:
	\begin{itemize}
		\item $\mathcal{C}_I$ acts on an object $(S,R)$ of $HSpa$ as $\mathcal{C}_I(S,R)=(HOM_{PSpa}((S,R),(\mathcal{L},\leq)),\rightarrow)$.
		\item $\mathcal{C}_I$ acts on an arrow $f:(S_1,R_1)\longrightarrow (S_2,R_2)$ in $HSpa$ as $\mathcal{C}_I(f):\mathcal{C}_I(S_2,R_2)\longrightarrow \mathcal{C}_I(S_1,R_1)$ defined by $\mathcal{C}_I(f)(\phi)=\phi\circ f$, where $\phi\in \mathcal{C}_I(S_2,R_2)$.

    \end{itemize}	
\end{defn}
\begin{lem}
	\label{D18}
	For each object $(S,R)$ of $HSpa$, $\mathcal{C}_I(S,R)$ is an object of $ISP_I(\mathcal{L})$.
	\begin{proof}
		By Note \ref{N1}, $HOM_{PSpa}((S,R),(\mathcal{L},\leq))$ is an object of $ISP(\mathcal{L})$. Thus, to prove $\mathcal{C}_I(S,R)$ as an object of $ISP_I(\mathcal{L})$, we shall show that if $f,g\in\mathcal{C}_I(S,R)$, then $f\rightarrow g\in \mathcal{C}_I(S,R)$. Now $(f\rightarrow g)^{-1}(\{\ell\})=\{s\in S:(f\rightarrow g)(s)=\ell\}$. By Definition \ref{D0}, we observe that $(f\rightarrow g)^{-1}(\{\ell\})=R^{-1}(g^{-1}(\{\ell\}))\cap (R^{-1}(f^{-1}(\{\ell\})))^c$. Now $R^{-1}(g^{-1}(\{\ell\}))\cap (R^{-1}(f^{-1}(\{\ell\})))^c$ is a clopen set in $S$, by the Definition \ref{D13}. Hence $f\rightarrow g\in\mathcal{C}_I(S,R)$, and then the intuitionistic implication operation$(\rightarrow)$ is well-defined. Therefore, $\mathcal{C}_I(S,R)$ is a subalgebra of intuitionistic power $\mathcal{L}^S$ of $\mathcal{L}$. So, $\mathcal{C}_I(S,R)$ is an object of $ISP_I(\mathcal{L})$.
	\end{proof}
\end{lem}
\begin{lem}
	\label{D19}
	For an arrow $f:(S_1,R_1)\longrightarrow (S_2,R_2)$ in $HSpa$, $\mathcal{C}_I(f)$ is an arrow in $ISP_I(\mathcal{L})$.
	\begin{proof}
		By Lemma \ref{D9}, $\mathcal{C}_I(f)$ is an arrow in $ISP(\mathcal{L})$. The only part we have to show that $\mathcal{C}_I(f)(g_1\rightarrow g_2)=\mathcal{C}_I(f)(g_1)\rightarrow \mathcal{C}_I(f)(g_2)$. Now for $s_1\in S_1$, $\mathcal{C}_I(f)(g_1\rightarrow g_2)(s_1)=(g_1\rightarrow g_2)\circ f(s_1)=\bigwedge\{g_1(y)\rightarrow g_2(y):f(s_1)R_2y\}$, and $(\mathcal{C}_I(f)(g_1)\rightarrow \mathcal{C}_I(f)(g_2))(s_1)=(g_1\circ f\rightarrow g_2\circ f)(s_1)=\bigwedge\{(g_1\circ f)(s_2)\rightarrow (g_2\circ f)(s_2):s_1R_1s_2\}=\bigwedge\{g_1(f(s_2))\rightarrow g_2(f(s_2)):s_1R_1s_2\}$. Since $f$ is order preserving, we observe that $\bigwedge\{g_1(y)\rightarrow g_2(y):f(s_1)R_2y\}\leq \bigwedge\{g_1(f(s_2))\rightarrow g_2(f(s_2)):s_1R_1s_2\}$. Again $f$ also satifies the condition given in the arrow part of Definition \ref{D13}, so we have $\bigwedge\{g_1(f(s_2))\rightarrow g_2(f(s_2)):s_1R_1s_2\}\leq \bigwedge\{g_1(y)\rightarrow g_2(y):f(s_1)R_2y\}$. Thus $\mathcal{C}_I(f)(g_1\rightarrow g_2)=\mathcal{C}_I(f)(g_1)\rightarrow \mathcal{C}_I(f)(g_2)$.
	\end{proof}
\end{lem}

Therefore, the functor $\mathcal{C}_I$ is well-defined by Lemma \ref{D18}, and Lemma \ref{D19}.

\begin{thm}
	\label{D20}
	For an object $(A,\rightarrow)$ of $ISP_I(\mathcal{L})$, $A$ is isomorphic to $\mathcal{C}_I\circ \mathcal{G}_I(A)$ in $ISP_I(\mathcal{L})$.
	\begin{proof}
	Define $\sigma_{(A,\rightarrow)}:A\longrightarrow \mathcal{C}_I\circ\mathcal{G}_I(A)$ by $\sigma_{(A,\rightarrow)}(a)(v)=v(a)$, where $a\in A$, and $v\in\mathcal{G}_I(A)$. It can be observed from Theorem \ref{D10}, that $\sigma_{(A,\rightarrow)}$ is an isomorphism in $ISP(\mathcal{L})$. So it is required to show that $\sigma_{(A,\rightarrow)}(a\rightarrow b)=\sigma_{(A,\rightarrow)}(a)\rightarrow \sigma_{(A,\rightarrow)}(b)$. Now $[\sigma_{(A,\rightarrow)}(a)\rightarrow \sigma_{(A,\rightarrow)}(b)](v)=\bigwedge\{\sigma_{(A,\rightarrow)}(a)(w)\rightarrow \sigma_{(A,\rightarrow)}(b)(w):vR_Aw\}=\bigwedge\{w(a)\rightarrow w(b):vR_Aw\}$, where $R_A$ is a partial order relation on $HOM_{ISP_I(\mathcal{L})}(A,\mathcal{L})$ defiened as in Definition \ref{D1}. From Definition \ref{D2}, it is observed that $\bigwedge\{w(a)\rightarrow w(b):vR_Aw\}=v(a\rightarrow b)=\sigma_{(A,\rightarrow)}(a\rightarrow b)(v)$. Hence, $\sigma_{(A,\rightarrow)}(a\rightarrow b)=\sigma_{(A,\rightarrow)}(a)\rightarrow \sigma_{(A,\rightarrow)}(b)$
	\end{proof}
\end{thm}
\begin{thm}
	\label{D21}
	For an object $(S,R)$ of $HSpa$, $(S,R)$ is isomorphic to $\mathcal{G}_I\circ\mathcal{C}_I(S,R)$ in the category $HSpa$.
	\begin{proof}
	Define $\delta_{(S,R)}:(S,R)\longrightarrow \mathcal{G}_I\circ\mathcal{C}_I(S,R)$ by $\delta_{(S,R)}(s)(f)=f(s)$, where $s\in S$, and $f\in\mathcal{C}_I(S,R)$. It is observed from Theorem \ref{D11} that $\delta_{(S,R)}$ is an isomorphism in the category $PSpa$. Also we observe in Theorem \ref{D11} that the relation $s_1Rs_2\iff\delta_S(s_1)R'\delta_S(s_2)$ hold, for any $s_1, s_2\in S$, and $R'$ is taken as the partial order relation on $\mathcal{G}\circ\mathcal{C}(S,R)$. So by definition of $\delta_{(S,R)}$ we can conclude that $\delta_{(S,R)}$ satisfies the relation $s_1Rs_2\iff\delta_{(S,R)}(s_1)R'\delta_{(S,R)}(s_2)$. Since $\delta_{(S,R)}$ is bijective and satifies the relation $s_1Rs_2\iff\delta_{(S,R)}(s_1)R'\delta_{(S,R)}(s_2)$, it can be easily shown that $\delta_{(S,R)}$ and $\delta^{-1}_{(S,R)}$ satisfy the condition given in the arrow part of Definition \ref{D13}.

	\end{proof}
\end{thm}

\begin{thm}
	\label{D22}
	The category $ISP_I(\mathcal{L})$ is dually equivalent to the category $HSpa$.
	\begin{proof}
	Let $ID_{ISP_I(\mathcal{L})}$, and $ID_{HSpa}$ denote the identity functors on $ISP_I(\mathcal{L})$ and $HSpa$, respectively. Define natural transformations $\sigma:ID_{ISP_I(\mathcal{L})}\longrightarrow \mathcal{C}_I\circ\mathcal{G}_I$, and  $\delta:ID_{HSpa}\longrightarrow\mathcal{G}_I\circ\mathcal{C}_I$. Then for each object $(A,\rightarrow)$ of $ISP_I(\mathcal{L})$, and $(S,R)$ of $HSpa$, we take $\sigma_{(A,\rightarrow)}:A\longrightarrow\mathcal{C}_I\circ\mathcal{G}_I(A)$ by $\sigma_{(A,\rightarrow)}(a)(v)=v(a)$, $v\in\mathcal{G}_I(A)$. Also for an object $(S,R)$ of $HSpa$, take $\delta_{(S,R)}:(S,R)\longrightarrow\mathcal{G}_I\circ\mathcal{C}_I$ by $\delta_{(S,R)}(s)(f)=f(s)$, $f\in\mathcal{C}_I(S,R)$. Now it can be easily checked that $\sigma$, and $\delta$ are indeed a natural transformations. By Theorem \ref{D20}, and Theorem \ref{D21}, $\sigma$, and $\delta$ are natural isomorphism. Therefore, the categories $ISP_I(\mathcal{L})^{op}$, and $HSpa$ are equivalent.
	\end{proof}
\end{thm}

\section{Conclusions:}
\label{sec:4}
We have shown a bitopological duality for algebras of Fitting's-logic by introducing the notion of the category $PBS_{\mathcal{L}}$. After that, we have proceed for developing a duality for the category $ISP_I(\mathcal{L})$. We defined intutionistic Kripke condition for $\mathbb{ISP_I}(\mathcal{L})$. Our first objective is to establish a duality for $ISP(\mathcal{L})$, where $\mathcal{L}$ is a finite bounded distributive lattice, and that has been shown using the category $PSpa$. As a result, the duality between the category $ISP_I(\mathcal{L})$ and the category $HSpa$ has been worked out.\\
Maruyama in \cite{maruyama2012natural}, develop a coalgebraic duality for the category $ISP_M(\mathcal{L})$ by define a suitable Vietoris functor. In this respect, we conceive that modalization of the notion of $\mathbb{ISP_I}(\mathcal{L})$ will yield a plausible result. A co-algebraic duality can be developed for the modalized notion of $\mathbb{ISP_I}(\mathcal{L})$, and that will be shown by defining a bi-Vietoris functor. We will attempt to develop this work at a later stage in the future.\\
By taking a suitable topological-systems, it would be possible to connect categorically with $ISP_I(\mathcal{L})$. So, another way to develop a duality for the defined category $ISP_I(\mathcal{L})$.

\end{document}